\documentclass[
prd,
nofootinbib,
superscriptaddress,
showpacs
]{revtex4}

\usepackage{graphicx}
\usepackage{epsfig}
\usepackage{latexsym}
\usepackage{amsmath}
\usepackage{amsfonts}
\usepackage{amsxtra}

\def\krto{ {\,\,\lower .8ex\hbox {$\longrightarrow \atop k \rightarrow 0$}\,\,}}
\def\Section#1{\section{#1}\hspace{\parindent}}

\def\bea{\begin{eqnarray} }
\def\beq{\begin{eqnarray} }

\def\eea{\end{eqnarray}}
\def\eeq{\end{eqnarray}}
\def\eq#1{eq.~(\ref{#1})}

\begin{document} 

\title{A brief comment on the similarities of the IR solutions for the ghost propagator 
DSE in Landau and Coulomb gauges}

\author{J. Rodr\'{\i}guez-Quintero}
\affiliation{Dpto. F\'isica Aplicada, Fac. Ciencias Experimentales; 
Universidad de Huelva, 21071 Huelva; Spain.}

\begin{abstract}

This brief note is devoted to reconcile the conclusions from a recent analysis 
of the IR solutions for the ghost propagator Dyson-Schwinger equations 
in Coulomb gauge with previous studies in Landau gauge.

\end{abstract}


\maketitle





\Section{Introduction}

In a series of papers~\cite{Boucaud:2008ji,Boucaud:2008ky, 
RodriguezQuintero:2010wy}, 
we studied the IR behaviour of the solutions for 
the ghost propagator Dyson-Schwinger equation (GPDSE) in Landau gauge. 
A similar analysis has been very recently carried out in Coulomb gauge~\cite{Watson:2010cn}. This brief 
note is devoted to reconcile the apparently discrepant pictures resulting from 
the studies in Landau and Coulomb gauges mentioned above.

\section{The GPDSE in Coulomb and Landau gauges}

As was explained in detail in 
refs.~\cite{Boucaud:2008ji,Boucaud:2008ky,
RodriguezQuintero:2010wy}, 
the low-momentum behavior for the Landau gauge ghost dressing function can be inferred from the 
analysis of the Dyson-Schwinger equation for the ghost
propagator (GPDSE) which can be written as follows:  
\beq
\frac{1}{F(k^2,\mu^2)} 
\ = \  
\frac{1}{F(p^2,\mu^2)}  \ + \ N_C \ g^2(\mu^2) \ H_1 \ I(k^2) \ ,
\label{SDRS}
\eeq
where $F(k^2,\mu^2)$ is the ghost propagator dressing function renormalized at 
the subtraction point, $\mu^2$, $H_1$ is the ghost-colinear non-perturbative 
ghost-gluon form factor (usually assumed to be 1) and 
\beq
\label{LamInf}
I(k^2) 
\ = \ 
\int \frac{d^4 q}{(2\pi)^4} 
\left( \rule[0cm]{0cm}{0.8cm}
\frac{F(q^2,\mu^2)}{q^2} \left(\frac{(k\cdot q)^2}{k^2}-q^2\right) 
\
\left[ \rule[0cm]{0cm}{0.6cm}
\frac{\Delta\left((q-k)^2,\mu^2\right)}{(q-k)^2} -  
\frac{\Delta\left((q-p)^2,\mu^2\right)}{(q-p)^2} 
\rule[0cm]{0cm}{0.6cm} \right]
\rule[0cm]{0cm}{0.8cm} \right) \ 
\eeq
with a renormalized gluon propagator, 
\beq\label{gluonprop}
\Delta(q^2,\mu^2) = \frac{B(\mu^2)}{q^2 + M^2} \ 
\simeq \frac{B(\mu^2)}{M^2} \left( 1 - \frac{q^2}{M^2} + \cdots \right) .
\eeq
where the Schwinger mechanism~\cite{Schwinger1962} is invoked to generate, via the fully dressed 
non-perturbative three-gluon vertex, a dynamical gluon mass, $M(q^2)$, that 
we will approximate by its zero-momentum value in the IR domain~\footnote{This is 
shown to be a good low-momentum approximation for the running mass in 
ref.~\cite{Aguilar:2009nf}.}. 

On the other hand, the authors of ref.~\cite{Watson:2010cn} recently performed a  
similar study of the GPDSE in Coulomb gauge. In particular, they applied the same 
strategy followed to investigate the low-momentum Landau-gauge ghost dressing function 
in ref.~\cite{Boucaud:2008ji} and took the Gribov's equal-time spatial gluon propagator 
dressing function~\footnote{In very good agreement with the Euclidean SU(2) lattice results 
obtained for small lattice couplings in ref.~\cite{Burgio:2008jr}.}, 
\beq\label{Gribov}
G^{T}(\vec{k}^2) = \int_{-\infty}^{\infty} \frac{dk_4}{2\pi} \ 
\frac{G\left(k_4^2,\vec{k}^2\right)}{k_4^2+\vec{k}^2} \ = \ 
\frac 1 2 \frac{\sqrt{\vec{k}^2}}{\sqrt{\vec{k}^4+m^4}} \ ,
\eeq 
as the input required to build a kernel and solve the GPDSE, again with the approximation 
of replacing the fully dressed spatial ghost-gluon vertex by the bare one (this is, also in 
Coulomb gauge, an exact result in the limit of a vanishing incoming ghost 
up to all perturbative orders~\cite{Watson:2006yq}). 
Thus, the GPDSE can be rewritten as follows:
\beq
\frac 1 {F\left(\vec{k}^2,\mu^2\right)} 
\ = \ 
\frac 1 {F\left(\vec{p}^2,\mu^2\right)}   
\ - \ 
N_C \frac{g^2(\mu)}{(4 \pi)^2} 
\int_{0}^{\infty} \frac{d\vec{q}^2}{\vec{q}^2} F(\vec{q}^2, \mu^2)  
\ 
\left( I\left(\vec{k}^2,\vec{q}^2; m \right) 
- I\left(\vec{p}^2,\vec{q}^2; m \right) \right) \ ,
\label{SDRSCou}
\eeq
where $I$ represents the angular integration, 
\beq
I\left(\vec{k}^2,\vec{q}^2; m \right) 
\ = \ 
\int_{-1}^{1} dz \left(1-z^2\right)  
\left(1+\frac {\vec{k}^2}{\vec{p}^2} 
- 2 z \sqrt{\frac {\vec{k}^2}{\vec{p}^2} } \right)^{-1/2}
\left[
\left(1+\frac {\vec{k}^2}{\vec{p}^2} 
- 2 z \sqrt{\frac {\vec{k}^2}{\vec{p}^2} } \right)^2 + \frac{m^4}{\vec{p}^4}
\right]^{-1/2}
\ .
\eeq
It should be emphasized that the ghost propagator dressing function in Coulomb 
gauge is strictly independent of the energy, $k_4^2$, as a non-perturbative result of the 
Slavnov-Taylor identities~\cite{Watson:2007vc}.

After assuming a pure powerlaw behaviour, $F(q^2) \sim (q^2)^{\alpha_F}$, for the ghost dressing 
function and analyzing asymptotically both Eqs.~(\ref{SDRS},\ref{SDRSCou}), one is left in both 
gauges with the two following well-known cases: (i) $\alpha_F=0$ (``{\it decoupling}''), 
that means zero-momentum finite ghost dressing function; and (ii) $\alpha_F \neq 0$ (``{\it scaling}''),  
where the low-momentum behavior of the gluon propagator, $\Delta(q^2) \sim (q^2)^{\alpha_G-1}$ 
forces the ghost dressing function to diverge at low-momentum through the scaling condition: 
$2 \alpha_F + \alpha_G = 0$. As well in Landau as in Coulomb gauge, a massive gluon propagator 
via the Schwinger mechanism or the Gribov's fomula for the equal-time spatial dressing 
lead to $\alpha_G=1$ and thus $\alpha_F=-1/2$. In particular, 
an ellaborated asymptotical analysis of \eq{SDRS} leaves us 
with~\cite{Boucaud:2008ky} :
\beq \label{solsFs}
F(q^2,\mu^2) \simeq 
\left\{
\begin{array}{lr}
\displaystyle
\left(
\frac {10 \pi^2}{N_C H_1 g_R(\mu^2) B(\mu^2)} 
\right)^{1/2}
\ \left(\frac {M^2} {q^2} \right)^{1/2} 
&
\mbox{\rm if } \alpha_F \neq 0 \ ,
\\
\rule[0cm]{0cm}{0.8cm}
\displaystyle
F(0,\mu^2) \left( 1 +   
\frac{N_C H_1}{16 \pi} \ \overline{\alpha}_T(0) \ 
\frac{q^2}{M^2} \left[ \ln{\frac{q^2}{M^2}} - \frac {11} 6 \right] 
\ + \ {\cal O}\left(\frac{q^4}{M^4} \right) \right)
&
\mbox{\rm if }  \alpha_F = 0 \ .
\end{array}
\right.
\eeq
If $\alpha_F \neq 0$, the perturbative strong coupling defined 
in the Taylor scheme~\cite{Boucaud:2008gn}, $\alpha_T=g_T^2/(4\pi)$, reaches a 
constant at zero-momentum,
\beq
\lim_{q^2 \to 0} \alpha_T(q^2) \ = \
\lim_{q^2\to 0} \left(
\frac{g^2(\mu^2)}{4 \pi}  q^2 \Delta(q^2,\mu^2) F^2(q^2,\mu^2) 
\right)
\ = \ 
\frac{5 \pi}{2 N_C H_1} \ ,
\label{alphaT0}
\eeq
as can be obtained from Eqs.(\ref{gluonprop},\ref{solsFs}).
In the case $\alpha_F = 0$, the subleading correction to the non-zero finite value for 
the zero-momentum ghost dressing function, given by \eq{solsFs},  
is controlled by the well-defined zero-momentum limit of $\overline{\alpha}_T(q^2)=(M^2/q^2) \alpha_T(q^2)$, 
which is an extension of the non-perturbative effective charge definition from the gluon  
propagator~\cite{Aguilar:2008fh} to the Taylor ghost-gluon coupling~\cite{Aguilar:2009nf}. 

The same two cases result from the analysis of \eq{SDRSCou} for the Coulomb gauge 
in ref.~\cite{Watson:2010cn}, where a ghost propagator dressing function 
behaving asymptotically as either a constant or $F(\vec{k}^2) \sim (\vec{k}^2)^{-1/2}$ is 
analytically found and confirmed by a numerical study. 

\section{The family of solutions and the dialing parameter}

\begin{figure}
\begin{center}
\begin{tabular}{cc}
\includegraphics[width=9cm,height=7cm]{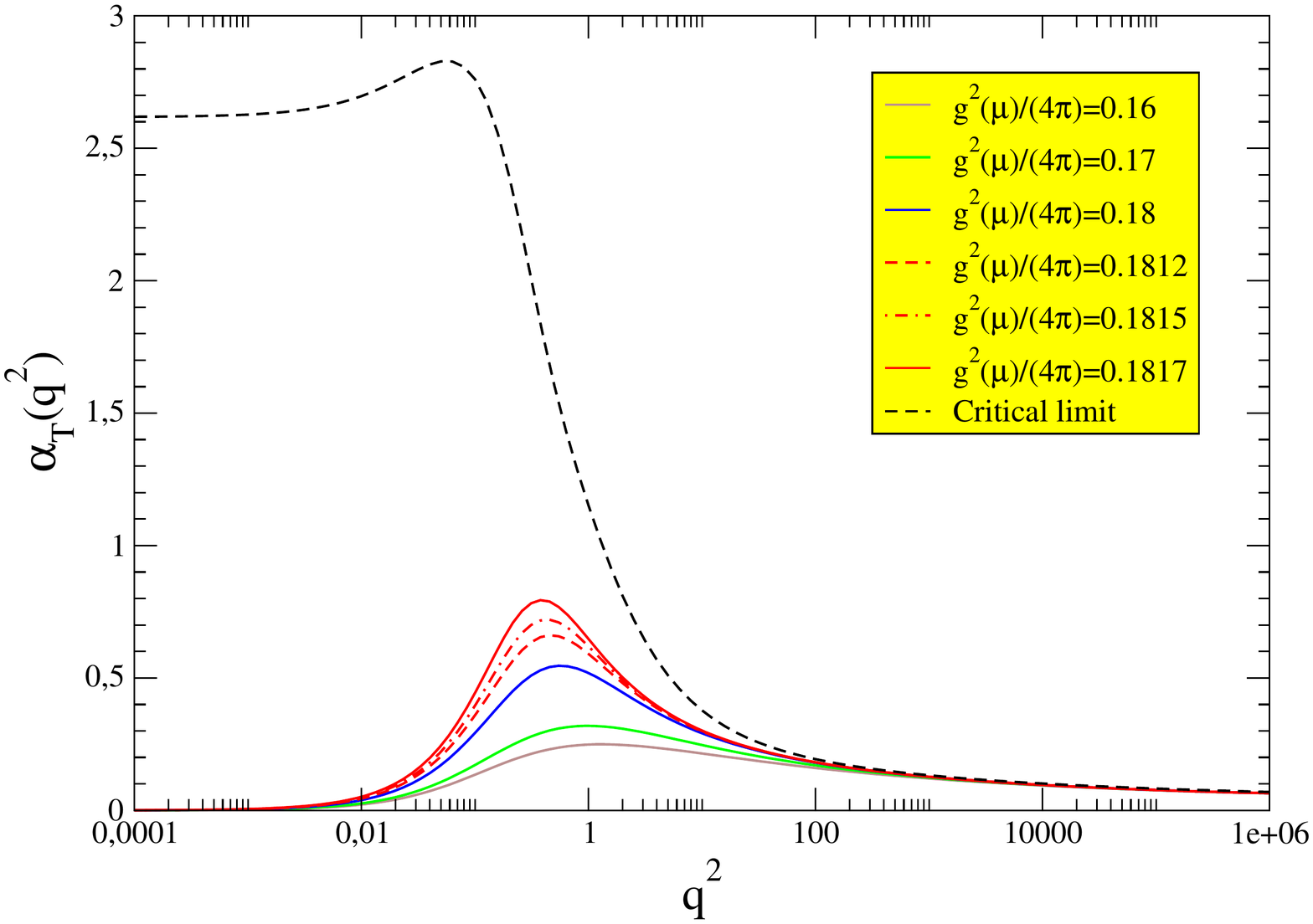} &
\includegraphics[width=9cm,height=7cm]{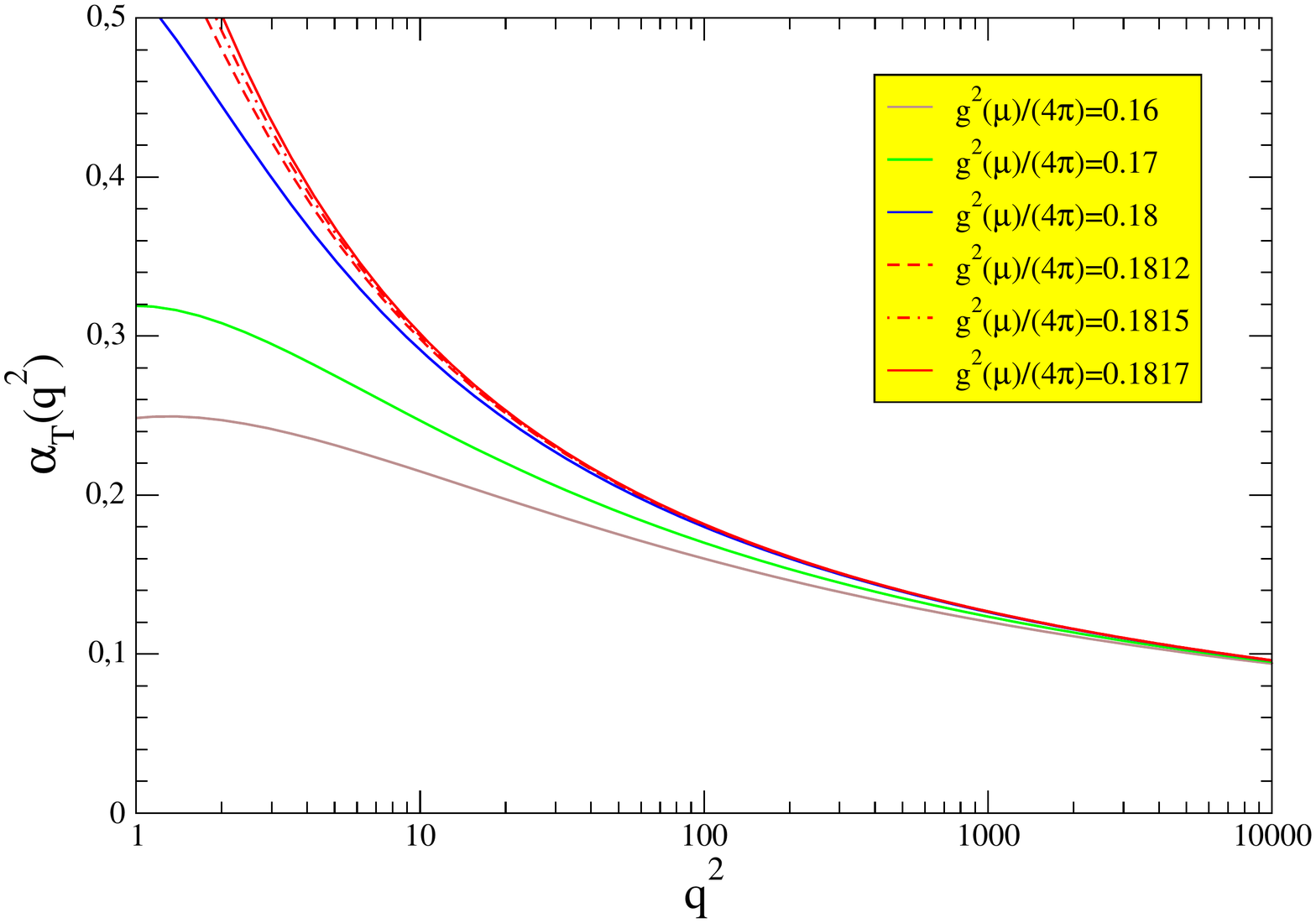} 
\end{tabular}
\end{center}
\vspace*{-0.75cm}
\caption{\small (Left) The Taylor coupling defined by \eq{alphaT0} and computed from the gluon and ghost propagator 
results of ref.~\cite{RodriguezQuintero:2010wy} for $g^2(\mu)/(4\pi)=0.16,0.17,0.18,0.1812,0.1815,0.1817$, with the 
subtraction point $\mu=10$ GeV. The curve for the critical limit is obtained by applying the 
results of ref.~ \cite{Boucaud:2008ji}, as explained in the text. (Right) The same in a ``zoomed'' 
high-momentum domain to show all the curves joining each other.}
\label{fig:alphaT}
\end{figure}

The GPDSE in \eq{SDRS} with the input of a gluon propagator borrowed from 
lattice QCD calculations is numerically solved in ref.~\cite{Boucaud:2008ji} 
and both kinds of solutions in \eq{solsFs} were shown to happen controlled 
by the size of the coupling at the renormalization point, $g(\mu)$. 
In QCD, one needs to provide a physical scale and a standard 
manner to proceed is by fixing the size of the coupling at a given momentum scale. 
This can be seen as a boundary condition to solve the DSEs. Thus, for any 
coupling, $g(\mu)$, below some critical value, $g_{\rm crit}$, an infinite number of 
regular or decoupling solutions for the ghost dressing, behaving as \eq{solsFs} 
indicates, were found; for $g(\mu)=g_{\rm crit}$, a unique 
critical or scaling solution behaving as \eq{solsFs} was found. It appeared not 
to be other solutions otherwise. In ref.~\cite{Boucaud:2008ji}, for a subtraction 
point $\mu=1.5$ GeV, a critical coupling $g_{\rm crit} \simeq 3.33$ and a very 
good description of ghost propagator lattice data with a regular solution of \eq{SDRS} for 
$g(\mu) \simeq 3.11$ were obtained. These results were also recently confirmed~\cite{RodriguezQuintero:2010wy} 
by studying 
the coupled system of ghost and gluon propagator DSE in the 
PT-BFM scheme~\cite{Binosi:2009qm}. This last work payed attention to 
the critical solution limit by studying how $F(0,\mu^2)$ diverges 
as $g(\mu) \to g_{\rm crit} \simeq 1.51$, with a subtraction point $\mu=10$ GeV. 
One can now apply the perturbative definition of the Taylor strong coupling in 
\eq{alphaT0} and compute this coupling with the gluon and ghost solutions of \cite{RodriguezQuintero:2010wy} 
in order to see how the critical limit is approached. This is shown in Fig.~\ref{fig:alphaT}, where 
it can be also seen that all the curves for $\alpha_T$ obtained for different values of $g(\mu)$ 
tend to join each other as $q^2/\mu^2$ increases (rigth). As explained 
in~\cite{RodriguezQuintero:2010wy}, the scaling solution cannot be obtained in the 
PT-BFM scheme with ``massive'' gluons but only appears as an end-point for the family 
of regular or decoupling solutions. In Fig.~\ref{fig:alphaT}, the curve for the critical 
limit is obtained by rescaling, up to giving $\alpha_T(0)$ from \eq{alphaT0} with $H_1=1$ at zero-momentum, 
the results at the critical limit for $q^2 \Delta(q^2) F^2(q^2)$ numerically obtained in ref.~\cite{Boucaud:2008ji},
not by solving the coupled DSE system but by applying the lattice gluon propagator as an 
input to solve the GPDSE. Indeed, the critical value for the coupling at $\mu=10$ GeV 
can be read from the critical curve in Fig.~\ref{fig:alphaT}  and one gets $g(\mu) \simeq 1.56$, 
in fairly good agreement with the value of ref.~\cite{RodriguezQuintero:2010wy}.

On the other hand, the numerical analysis of \eq{SDRSCou} for the Coulomb gauge 
in ref.~\cite{Watson:2010cn} also shows both regular and critical solution to happen 
but controlled by $F(0,\mu)$ (or $\Gamma(0,\mu)=1/F(0,\mu)$) as a boundary condition with 
the size of the coupling fixed to be $g^2(\mu)=\overline{g}^2=4 \pi \times 0.1187$ 
for $N_C=3$. In that case, regular or decoupling 
solutions correspond to finite values of $F(0,\mu^2)$ and the divergent limit, 
$F(0,\mu^2) \to \infty$, provides us with the unique critical or scaling solution (a similar 
pattern is claimed to be also found in Landau gauge for the authors of 
ref.~\cite{Fischer:2008uz}). Again, the lattice results for the ghost propagatot seem 
to agree with a regular solution with $F(0,\mu^2) \simeq 10$, although much larger values 
for the boundary condition cannot be ruled out. 
Furthermore, the authors of ref.~\cite{Watson:2010cn} demonstrate that 
the ghost dressing obtained from \eq{SDRSCou} for any $F(0,\mu)$ behave 
as $c \cdot (\vec{k}^2)^{\gamma_g}$ 
at asymptotically large $\vec{k}^2/m^2$; where $m$ is the Gribov mass scale in \eq{Gribov}, 
$\gamma_g$ is the leading-order ghost-anomalous dimension and $c$ is the same coefficient 
for all the solutions. In other words, in the perturbative domain, the perturbative behaviour 
is recovered and the differences between values of $F(\vec{k}^2,\mu^2)$ for 
different arbitrary inputs of $F(0,\mu^2)$ vanish (This can be clearly seen in 
Fig.~2 of \cite{Watson:2010cn}). Then, they conclude that the boundary condition 
is not connected to the renormalization (at least in the perturbative regime).

We are thus left with either a family of Landau-gauge DSE solutions dialed by the size of the coupling 
at the renormalization point or a family of Coulomb-gauge ones dialed by the zero-momentum 
ghost dressing value as a boundary condition not connected to the renormalization. How both pictures 
can be reconciled? The key point stems from the different renormalization prescriptions applied 
to the ghost propagator in both analyses. In the Landau-gauge analysis of 
refs.~\cite{Boucaud:2008ji,Boucaud:2008ky, 
RodriguezQuintero:2010wy}, the 
standard MOM prescription, where the Green functions are required to take their 
tree-level expression at the renormalization point and for some particular kinematical choice 
(this implies $F(\mu^2,\mu^2)=1$), is the one applied. The prescription applied to the ghost propagator 
by the authors of ref.~\cite{Watson:2010cn} is defined by their eq.~(3.20) for  
the renormalization constant $Z_c(\Lambda,[\overline{g},\Gamma(0)])$, where $\Gamma(0)=1/F(0,\mu^2)$.
In particular, this renormalization constant depends on the boundary condition, $\Gamma(0)$, 
in such a manner that the value for this boundary condition is rescaling the ghost 
dressing function (and, as can be clearly seen in Fig.~2 of \cite{Watson:2010cn}, it does not 
take the tree-level value, 1, as happens in MOM prescription for the subtraction point). 
In the following, we will show how, depending on the 
renormalization prescription, both patterns can be found for the family 
of solutions for \eq{SDRS} in Landau gauge. 

Let's consider a MOM solution of \eq{SDRS} for arbitrary coupling, $g(\mu)$; let's then 
apply the following transformation:
\beq\label{MOM2New}
g(\mu) \to s \ g(\mu) \ , \quad F(q^2,\mu^2) \to \frac 1 s F(q^2,\mu^2) \ .
\eeq
The properties of \eq{SDRS} (the same happens of course for \eq{SDRSCou}) guarantee that, for 
any $s$ being a c-number, the transformed dressing function verifies the DSE equation with the 
transformed coupling (of course, MOM prescription implies $s=1$). 
Then, if one chooses $s=\overline{g}/g(\mu)$ and apply the transformation 
to every solution of the MOM family, we will be left with a one-to-one correspondence 
between these solutions and the new ones
\beq
\overline{F}(q^2,\mu^2)\ \equiv \ \frac {g(\mu)}{\overline{g}} F(q^2,\mu^2) \ ,
\eeq
for the fixed coupling $\overline{g}$, which can be identified by 
the zero-momentum value, $\overline{F}(0,\mu^2)$. This new family of transformed 
solutions obeys the same pattern of the Coulomb gauge family in ref.~\cite{Watson:2010cn}. 
It is interesting to notice that the strong coupling 
defined in the Taylor scheme can be also obtained from the transformed solutions as
\beq\label{TralphaT}
\alpha_T(q^2) \ \equiv \frac{\overline{g}^2}{4 \pi} \ q^2 \Delta(q^2,\mu^2) \overline{F}(q^2,\mu^2) 
\ \equiv \ \frac{g^2(\mu)}{4 \pi} \ q^2 \Delta(q^2,\mu^2) F(q^2,\mu^2) \ ,
\eeq
although it is obvious that neither $\overline{F}$ nor the coupling are in MOM scheme. 
More interesting enough is to realize from \eq{TralphaT} that: 
(i) $\overline{F}$ does not depend on $\mu$, as far as one applies the 
same renormalized gluon propagator to obtain any solution 
for arbitrary coupling, $g(\mu)$, as done in \cite{Watson:2010cn}; 
(ii) the transformed ghost dressing function for the critical MOM solution ($g(\mu) \to g_{\rm crit}$) 
corresponds to the scaling solution for the critical boundary condition, 
$\overline{F}(0,\mu^2) \to \infty$ and diverges as $\overline{F}(q^2,\mu^2) \sim (q^2)^{-1/2}$ 
for $q^2 \to 0$, as it clearly results from \eq{alphaT0}; 
(iii) finally, as the difference between the Taylor strong couplings computed 
for any two arbitrary values of $g(\mu)$ vanishes at asymptotically large momentum (see 
Fig.~\ref{fig:alphaT}), the same should happen to $\overline{F}$ for different values 
of $\overline{F}(0,\mu^2)$. 
All this is  claimed by the authors of \cite{Watson:2010cn} to identify 
the family of solutions for the Coulomb-gauge \eq{SDRSCou}. 

\section{Conclusions}

We thus conclude that the behaviour of the family of Coulomb-gauge GPDSE solutions in ref.~\cite{Watson:2010cn} 
is analogous to the one described in refs.~\cite{Boucaud:2008ji,Boucaud:2008ky, 
RodriguezQuintero:2010wy}
for Landau gauge, although not renormalized in MOM scheme but after 
applying the transformation of \eq{MOM2New}. The input parameter for the solutions in ref.~\cite{Watson:2010cn} 
is the zero-momentum ghost dressing, which can be interpreted as a boundary condition and put in connection 
with the Gribov problem. On the other hand, for Landau gauge and MOM scheme, $g(\mu)$ is related to the strong coupling 
in Taylor scheme while the size of the fixed coupling, $\overline{g}$, after applying \eq{MOM2New} is physically meaningless. 
Thus, as done in ref.~\cite{RodriguezQuintero:2010wy}, the critical value for $g(\mu)$ can be used to derive a critical value 
for $\Lambda_{\overline{\rm MS}}$ which can be compared with lattice evaluations or experimental determinations to 
investigate whether the critical solution can be ruled out.



%
%
\end{document}